\documentclass[a4paper,11pt]{article}
\pdfoutput=1 

\usepackage{jinstpub} 

\title{TCAD simulations of pixel sensors for the ATLAS ITk upgrade and performance of annealed planar pixel modules}

\author[1]{J.-C.~Beyer,\note{Corresponding author.}}
\author{A.~La~Rosa,}
\author{A.~Macchiolo,}
\author{R.~Nisius,}
\author{N.~Savic}
\author{and R.~Taibah}

\affiliation{Max-Planck-Institut f{\"u}r Physik (Werner-Heisenberg-Institut), \\ F{\"o}hringer Ring 6, DE-80805 M{\"u}nchen, Germany}

\emailAdd{jbeyer@mpp.mpg.de}

\abstract{For the high luminosity phase of the Large Hadron Collider to start operation around 2026, a major upgrade of the ATLAS Inner Tracker (ITk) is in preparation. Thanks to their low power dissipation and high charge-collection efficiency after irradiation, thin planar pixel modules are the baseline option to instrument all, except for the innermost layer of the pixel detector.
\\
To optimise the sensor layout for a pixel cell size of $50\times50\,\mu m^2$, TCAD simulations are being performed. Charge-collection efficiency, electronic noise and electrical-field properties are investigated. A radiation-damage model is employed in TCAD simulations to estimate the performance before- and after irradiation.
\\
The impact of storage time at room temperature for the ITk pixel detector during maintenance periods are estimated using sensors irradiated up to a fluence of 5$\times10^{15}\,$n$_\text{eq}$/cm$^2$. Pixel sensors of $100-150\,\mu m$ thickness, interconnected to FE-I4 read-out chips with pixel dimensions of $50\times250\,\mu m^2$, are characterised using the testbeam facilities at the CERN-SPS and DESY. The charge-collection and hit efficiencies are compared before and after annealing at room temperature for up to one year.}


\proceeding{20$^{\text{th}}$ International Workshop on Radiation Imaging Detectors\\
  24$^{\text{th}}$-28$^{\text{th}}$.06.2018\\
  Sundsvall, Sweden}

\begin{document}
\notoc
\maketitle
\flushbottom

\section{Introduction}

The Large Hadron Collider (LHC) is upgraded to achieve an increase of the instantaneous luminosity to about a factor of five. The accompanying increase in simultaneous proton-proton interactions within one bunch collision (pile-up events), will increase the track density in the tracking system. Employing smaller pixel cells is essential to keep the present occupancy and tracking resolution unchanged. Ten years of running with an instantaneous luminosity of $5-7\times10^{34}\,$cm$^{-2}$s$^{-1}$ will result in $4000\,$fb$^{-1}$ leading to a particle fluence of $2.6\times10^{16}\,$n$_\text{eq}/$cm$^2$ at the innermost parts of the experiment \cite{BSmart,StripsTDR}.

The ATLAS detector will be equipped with a new Inner Tracker (ITk) to cope with the harsher conditions. The ITk features an all silicon layout consisting of 5 pixel layers. To maintain the current occupancy, the size of the pixel cell is reduced from the present $50\times400\,\mu\mathrm{m}^2$ (layer 1-3, assembled with the FE-I3 readout chip) and $50\times250\,\mu\mathrm{m}^2$ (layer 0, Insertable B-Layer, assembled with the FE-I4 readout chip \cite{FEI4}) to $50\times50\,\mu\mathrm{m}^2$ or $25\times100\,\mu\mathrm{m}^2$. The RD53 collaboration is currently developing the necessary new read-out chip in the TSMC $65\,$nm CMOS technology \cite{RD53}. Due to the harsh radiation environment in the region close to the interaction point an exchange of the two innermost layers is foreseen after half of the detector lifetime. The replacement of these two layers may cause an interruption in the operation of the pixel detector cooling system. At this point in time, the third layer will have received a fluence in the range of $1-2\times10^{15}\,$n$_\text{eq}/$cm$^2$ and will undergo a reverse annealing, possibly leading to an increase in full depletion voltage and decrease in charge collection. To assess these effects, a measurement of the efficiency of annealed irradiated pixel modules is presented based on testbeam data.

This paper presents results for a Technology Computer Aided Design (TCAD) device simulation of thin planar pixel sensors, the baseline option to instrument all except for the innermost pixel layers, given their low material budget, high charge-collection efficiency after irradiation and low power dissipation. The simulation is carried out to optimise the pixel implant width within the $50\times50\,\mu\mathrm{m}^2$ pixel cells. Three observables were studied in the context of this investigation: breakdown voltage, pixel capacitance and charge collection efficiency (CCE).

\section{Thin planar pixel sensors}
\label{thinplanarpixelsensors}

Experimentally, thin planar pixel sensors with a thickness of $100\,\mu\mathrm{m}$ have been shown to efficiently perform after fluences up to $1\times10^{16}\,$n$_\text{eq}/$cm$^2$, fulfilling the requirements to instrument layer 1-4 of the ITk \cite{NSavic}. The moderate bias voltage necessary to obtain a tracking efficiency $\geq97\%$ and the corresponding lower power dissipation compared to thicker sensors make them particularly suited for the application at the HL-LHC fluences. They are based on the n-in-p technology with a homogeneous maskless low-dose p-spray boron implantation ensuring inter-pixel isolation. Investigations are performed with simulations and measurements on a sensor production carried out by the Semiconductor Laboratory (HLL) of the Max-Planck Society. To process very thin substrates, 6 inch Silicon on Insulator (SOI) wafers were used with $100\,\mu\mathrm{m}$ and $150\,\mu\mathrm{m}$ active thicknesses. The technology foresees the removal of the handle wafer after the post-processing phase in which the Under Bump Metal (UBM) is deposited.

\section{Simulation methods and setup}


For all simulation results presented, the finite-element semiconductor simulation package Synopsys Sentaurus TCAD version L-2016.03 was used. A 3D model of the pixel sensor consists of a mesh structure of individual vertices. To reduce the computation time, while retaining the predictive power, the structure can be reduced to its smallest unit, in this case four quarter pixels. Multiple pixels need to be considered to enable the program to compute the weighting field essential to simulate correctly the charge collection properties. The repetition of the structure is dealt with Neumann boundary conditions requiring the electrical fields and electrical currents to be zero normal to the boundaries of the mesh. This requirement and hence, the high accuracy of the unit is given if symmetry-planes are chosen as boundaries. It was shown \cite{Rainer} that this assumption is valid for clean wafers before irradiation.
\\
A broad range of physics models is enabled in the program to ensure realistic conditions, see Ref. \cite{SynopsysManual}. For the simulation of charge carrier mobility the doping dependence is described by the Masetti model, while the high-field saturation of the mobility is described by the extended Canali model, using the gradient of the Quasi-Fermi potential as driving force. The recombination of charge carriers follows temperature and doping dependent Shokley-Read-Hall statistics, also considering Hurx tunneling. The van-Overstraeten and de-Man model was chosen to describe charge multiplication due to avalanche processes. The bandgap narrowing is reproduced by the Old-Slotboom model. Finally, mobility degradation at interfaces is computed according to the enhanced Lombardi model.
\\
The simulated volume is $50\times50\times\mathrm{thickness}\,\mu\mathrm{m}^3$. It includes the insulation layers of high temperature oxide, nitride and Low Temperature Oxide (LTO) on top of the silicon bulk, in correspondence to the actual process of the manufactured sensors. An electrode on top of the pixel mimics the aluminium contact. The size of the metal overlap depends on the width of the implant. A fixed overlap of $2\,\mu m$ was chosen for the investigation of the implant width on the breakdown voltage and capacitance in Sec. \ref{breakdown} and Sec. \ref{sec:capacitance} to disentangle the effects of implant width and metal overhang. For charge collection simulations, an overlap of $6\,\mu$m was chosen for an implant width of $10\,\mu$m, $4\,\mu$m overlap for $20\,\mu$m and $30\,\mu$m implant width and $2.5\,\mu m$ for $40\,\mu$m implant width. The connection of the electrode to the pixel implant is realised through a $5\times5\,\mu\mathrm{m}^2$ via in the insulation layers. The bulk doping is about $10^{12}\,$cm$^{-3}$ and the p-spray has a peak concentration of $6\times10^{16}\,$cm$^{-3}$ while it reaches a depth of $0.6\,\mu\mathrm{m}$ until the concentration vanishes. The backside of the sensor has a homogeneous p$^+$ doping with and an electrical contact. The bias voltage is applied to the sensor by keeping the pixel contact at ground while ramping the potential of the backside contact to (negative) high voltage.
\\
One important goal is to investigate the consequences of irradiation depending on the sensor design. Irradiation causes two main categories of radiation damages: bulk damages and surface effects. Bulk damage manifests as an increase of leakage current, full depletion voltage and trapping probability leading to a decreased carrier lifetime. This can be explained by the non-ionising-energy-loss (NIEL) hypothesis. Particles of high energy traverse the silicon sensor and loose energy by ionisation and scattering with bulk atoms. In case of significant energy transfer, the bulk atom is displaced from its original lattice position causing a vacancy. Depending on the actual scattering, the energy level of the vacancy can either be of acceptor or donor type, but will always be situated in the band-gap of silicon. The acceptor type vacancies can trap electrons and thereby explain the increase of full depletion voltage (filled traps are negatively charged and increase the effective p-type doping) and increased trapping. Since the vacancies also act as carrier generation centers, they also cause the increased leakage current. In Synopsys TCAD, bulk damage is simulated by the introduction of acceptor and donor type traps. The model is taken from Ref. \cite{newPerugia} and contains two acceptor and one donor type traps. This model was chosen as one amongst several available models as it is being widely used in the community. While the absolute differences between the available models might be significant, the comparison of the relative performance of different pixel geometries are likely much less affected.
\\
The second category of radiation damage are surface effects. Ionising radiation releases electron hole pairs in the insulating SiO$_2$ layer grown on top of the bulk. The electrons have a sufficient mobility. Holes can only be transported via shallow traps in the oxide, and are captured by deep traps forming a fixed positive space charge. Following the measurements of Ref. \cite{Zhang2012}, the oxide charge saturates after high ionising doses in the order of several hundred MGy at values of $\approx2-3\times10^{12}$cm$^{-2}$. Due to the spread arising from different production technologies and given the possibly not completely saturated and already partially annealed oxide charge, the oxide charge in the simulation is fixed to $1.5\times10^{12}\,$cm$^{-2}$ for all irradiations and to $5\times10^{10}\,$cm$^{-2}$ before irradiation \cite{RainerPrivate}.
\\
The temperature is $300\,$K for all not-irradiated devices and $254\,$K for all irradiated devices. The charge collection efficiencies are obtained by simulating the impact of a MIP releasing 76 electron/hole pairs per $\mu\mathrm{m}$ in a 100$\,\mu$m thick sensor (the value is higher for thicker sensors) with a Gaussian width of $1\,\mu\mathrm{m}$. The transient of the device is simulated from $5\,$ns before the impact to $30\,$ns after the impact. The collected charge is obtained by integrating the current pulse and subtracting the baseline leakage current which is extracted from the $5\,$ns before the impact.

\section{Simulation of \texorpdfstring{\boldmath{$50\times50\,\mu\mathrm{m}^2$}}{50x50um2} pixel cells}

\subsection{Investigation of breakdown behaviour}
\label{breakdown}

A breakdown of the sensor occurs if the electrical field is sufficient to accelerate charge carriers on their mean free path enough to ionise secondary charge carriers in the scattering process. This avalanche process results in an excessive current which makes the device inoperable. The region of highest electrical field is defined by the highest gradient of the doping concentration in case of a pn-junction. This occurs at the interface of p-spray and n$^+$ implant since the p-spray concentration is usually several orders of magnitude larger than the doping of the high resistive bulk material. The potential difference between pixel implant and surrounding p-spray defines the height of the electrical field. The potential of the implant is fixed to ground through the readout channel, whereas the p-spray adjusts its potential between backside potential and ground. Here, the size of the pixel plays an important role as it shields the p-spray from the backside potential, hence, holding it closer to ground the larger the implant is.

\begin{figure}[tbph!]
\centering
\includegraphics[trim=0.5cm .5cm .3cm 1.1cm, clip,height=4.5cm]{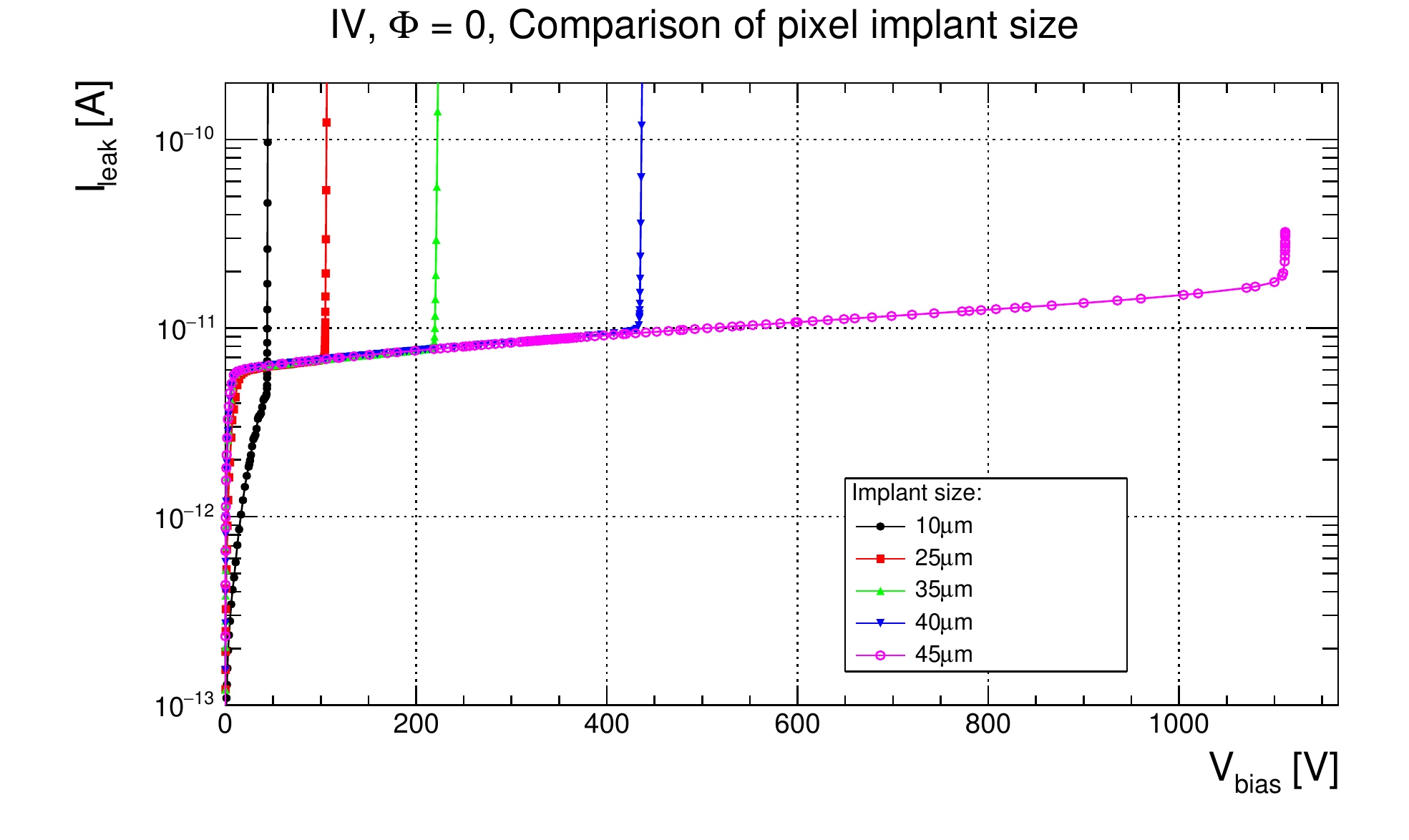}
\includegraphics[height=4.5cm]{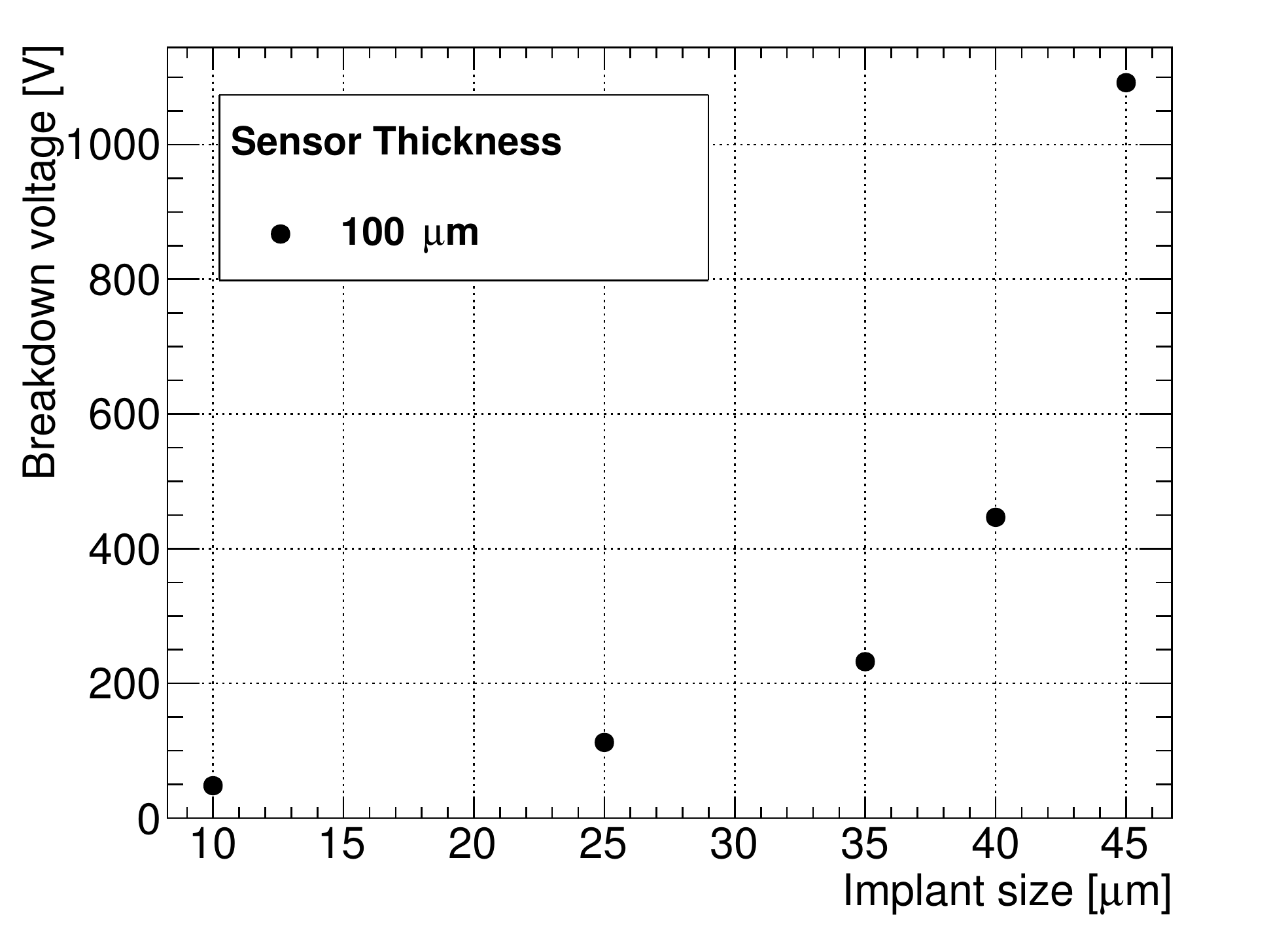}
\caption{\label{fig:bdvsimplantsize} Simulated current of pixels of different sizes as a function of applied bias voltage for 100$\,\mu$m thick sensors (left). Breakdown voltage as a function of the implant width (right).}
\end{figure}

Figure \ref{fig:bdvsimplantsize} shows the simulated breakdown voltage of the pixel cell in the 3D model. In agreement with the expectations, the breakdown voltage increases sizeably for increasing implant width. The shielding properties of the implant regarding the p-spray layer enables the much higher breakdown voltages observed for larger implants. Test-structures with different pixel implant widths are implemented in a new production of SOI wafers at HLL to experimentally validate the results presented in this study.

\subsection{Pixel capacitance}
\label{sec:capacitance}
The input capacitance seen by the pre-amplifier of the readout chip is the main driver of the noise. Operation of the ITk pixel detector at very low thresholds will be essential to maintain the required hit efficiency after the high fluences expected at the HL-LHC. Consequently, the noise has to be as low as possible. Therefore, the capacitance of the implant is one important observable in the overall optimisation of the implant width.

\begin{figure}[tbph!]
\centering
\includegraphics[trim=0.8cm .6cm .3cm 1.1cm, clip, width=0.49\textwidth]{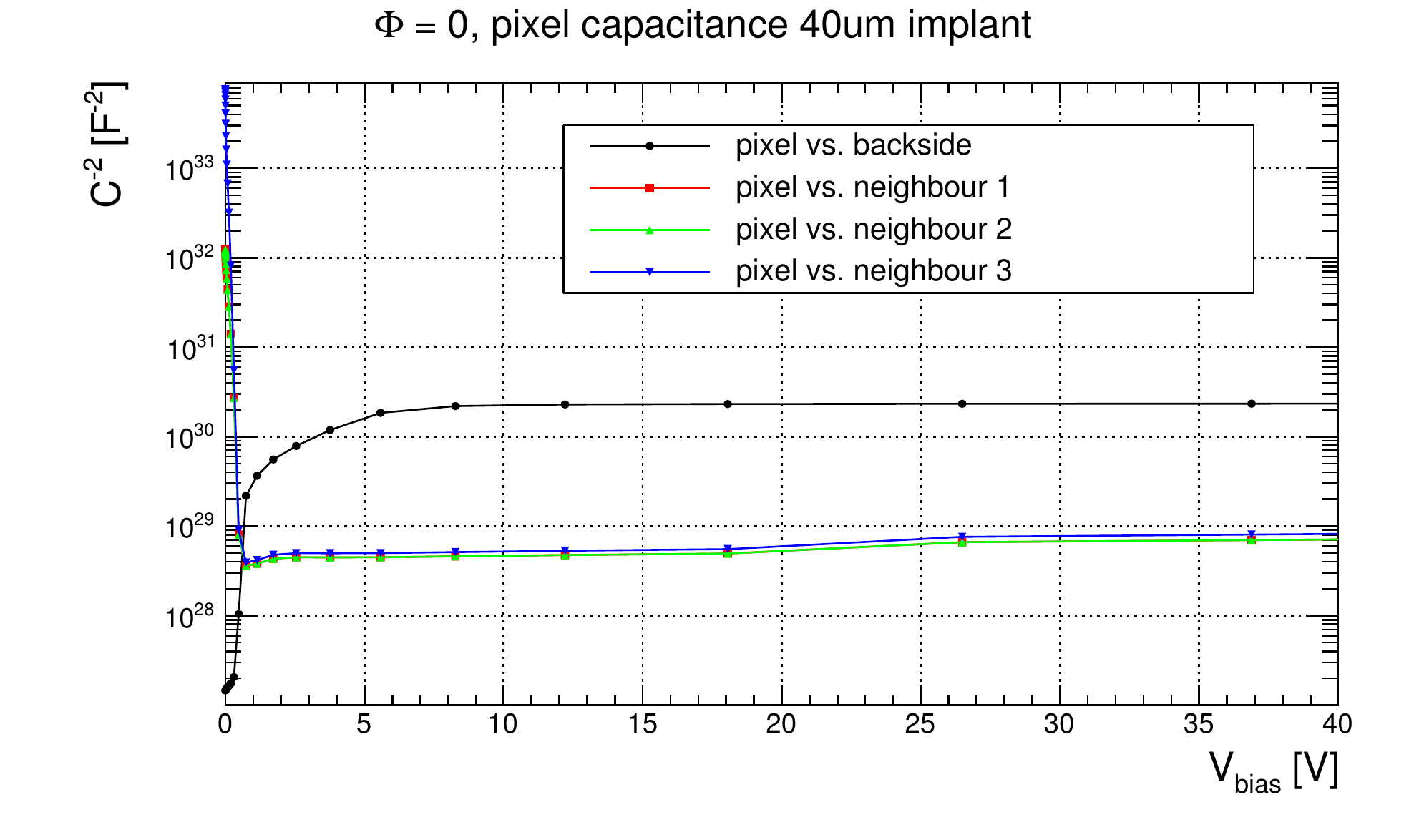}
\includegraphics[trim=0.8cm .6cm .3cm 1.1cm, clip, width=0.49\textwidth]{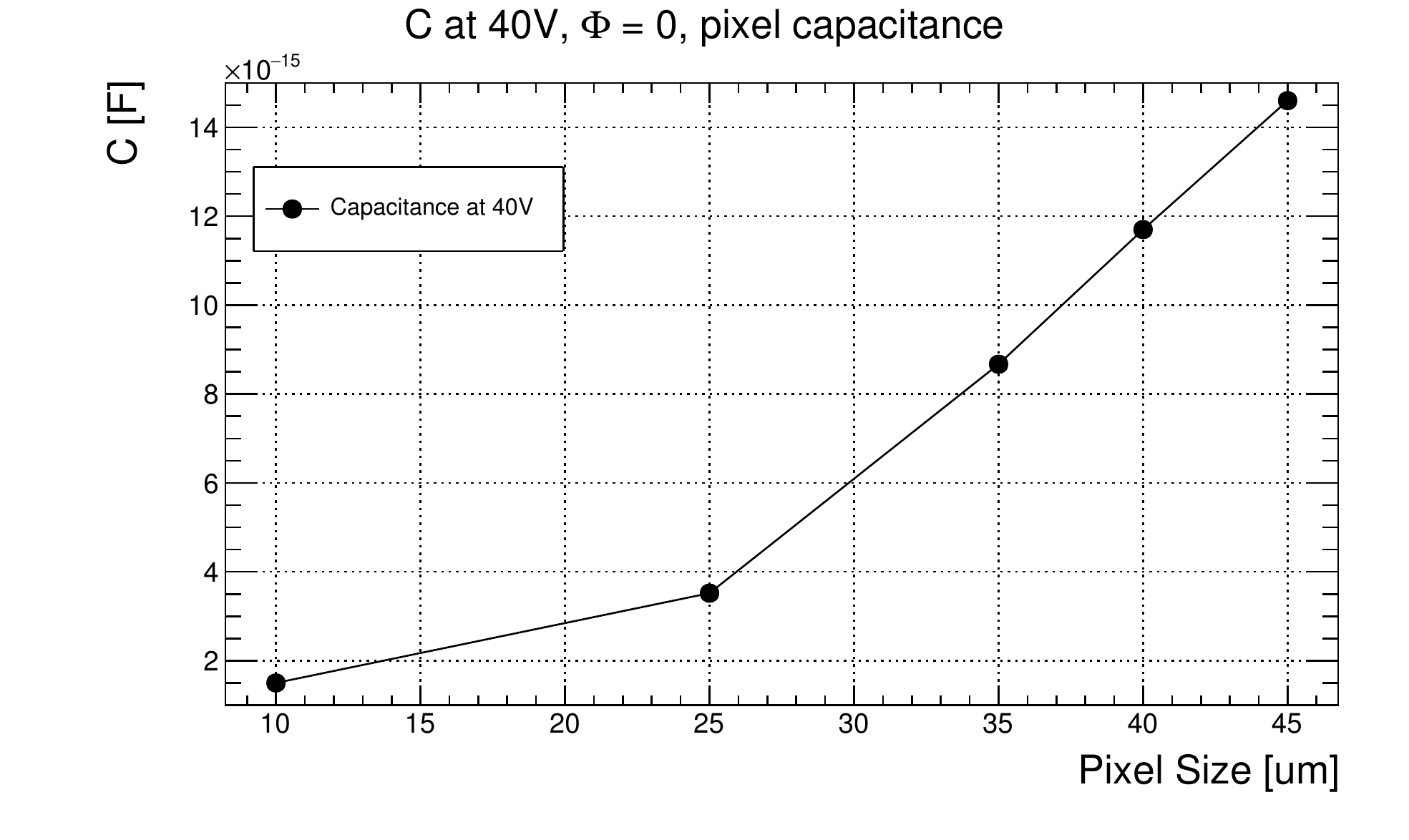}
\caption{\label{fig:capacitancevsimplantsize} Squared inverse capacitance of the different capacitance contributions as functions of the bias voltage for an implant width of $40\,\mu m$ at a frequency of $10\,$kHz (left). The total capacitance is shown as a function of the implant width for a quarter pixel and for a bis voltage of $40\,$V (right).}
\end{figure}

Figure \ref{fig:capacitancevsimplantsize} displays the inverse squared capacitance as a function of the bias voltage on the left. The applied frequency is $10\,$kHz and an implant width of $40\,\mu$m was chosen as an example to illustrate the composition of the total capacitance. The inter-pixel capacitance by far outweighs the capacitance to the backside once the sensor is depleted. The capacitance to the diagonal neighbour (3) is slightly smaller than the one to the direct neighbours (1 and 2, overlaid). Figure \ref{fig:capacitancevsimplantsize} (right) shows the total capacitance for one of the quarter pixels. For example, an implant width of $35\,\mu$m results in a total pixel capacitance of $36\,$fF, thus, being well below the $50\,$fF limit up to which the minimal operable threshold of the RD53 readout chip is guaranteed.

\subsection{Charge collection efficiency}

As another important benchmark, the influence of the pixel implant width on the charge-collection efficiency (CCE) is also investigated. The CCE is calculated as the fraction of the charge that is being collected by the largest implant (40$\,\mu$m) at $40\,$V before irradiation. Fluences of 0, 1, 3 and $5\times10^{15}\,$n$_\text{eq}/$cm$^2$ were investigated. The collected charge is evaluated at a fixed position (10$\,\mu$m/10$\,\mu$m, 0/0 being the center of the four pixel region) while varying the bias voltage. From this, a voltage point before and after saturation of collected charge is chosen. For these points, the spatial dependence of the collected charge is studied at the previously determined voltage points by changing the penetration point across the diagonal of the collecting pixel.

\subsubsection{Dependence of CCE on bias voltage}
\label{sec:ccevsvoltage}

\begin{figure}[tbph!]
\centering
\includegraphics[width=0.8\textwidth]{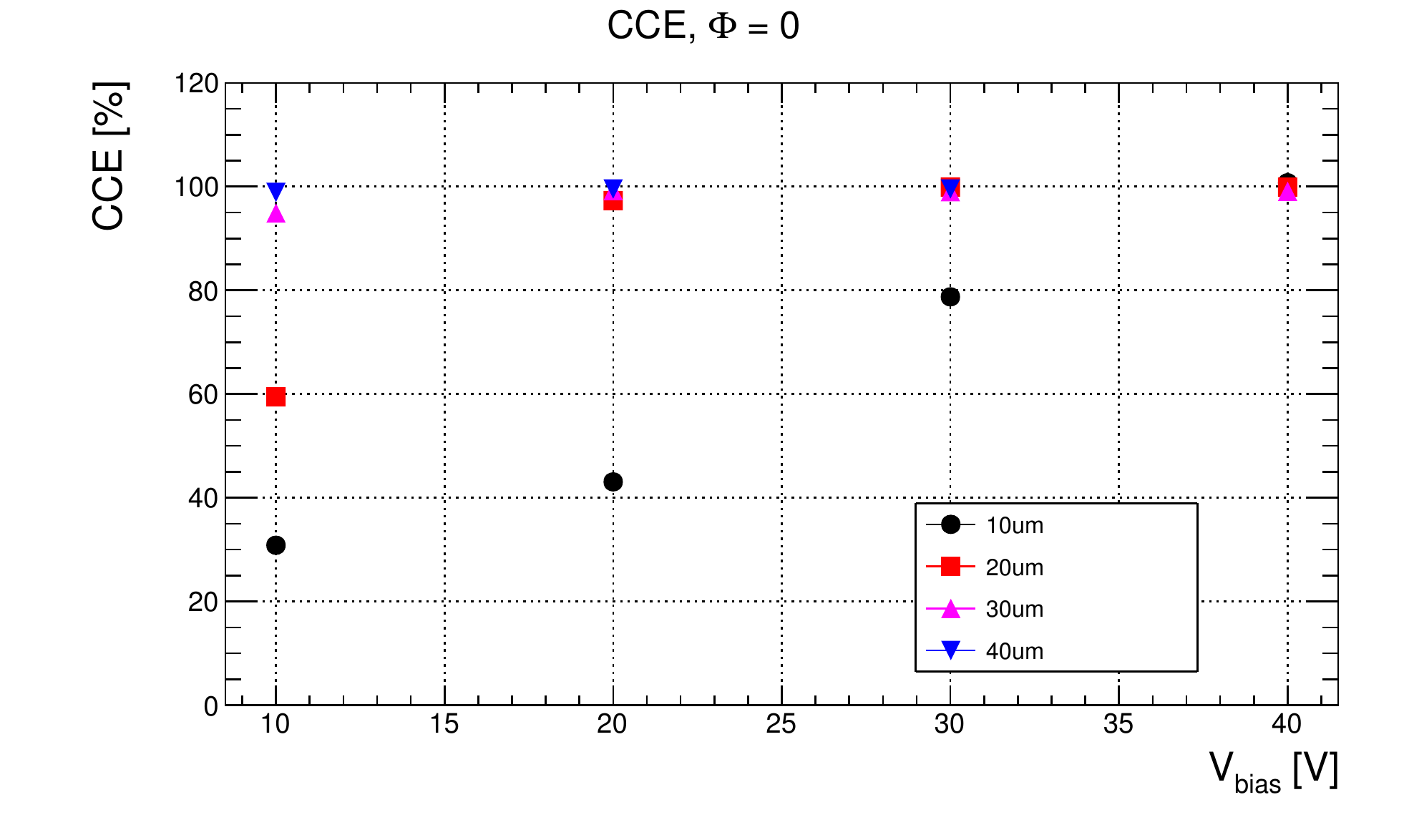}
\caption{\label{fig:CCEvsbiasfl0} Charge collection efficiency for different implant widths as a function of bias voltage before irradiation. The particle traverses the sensor perpendicularly at 10$\,\mu$m/10$\,\mu$m, with 0/0 being the center of the four pixel region.}
\end{figure}

The CCE increases with bias voltage up to a saturation value for all implant widths at all stages of irradiation. Figure \ref{fig:CCEvsbiasfl0} shows the behaviour before irradiation. The largest implant ($40\,\mu$m) shows full efficiency already at the lowest tested voltage of $10\,$V. Instead, smaller implants tend to deplete later and consequently, they show a later saturation of the charge collection efficiency. The $10\,\mu$m wide implant achieves full collection efficiency only at $40\,$V. Thus, $10\,$V and $40\,$V will be the two bias voltage values investigated before irradiation in Sec. \ref{section:spatialCCE}.

\begin{figure}[tbph!]
\centering
\includegraphics[trim=1.2cm .6cm .3cm 1.1cm, clip, width=0.49\textwidth]{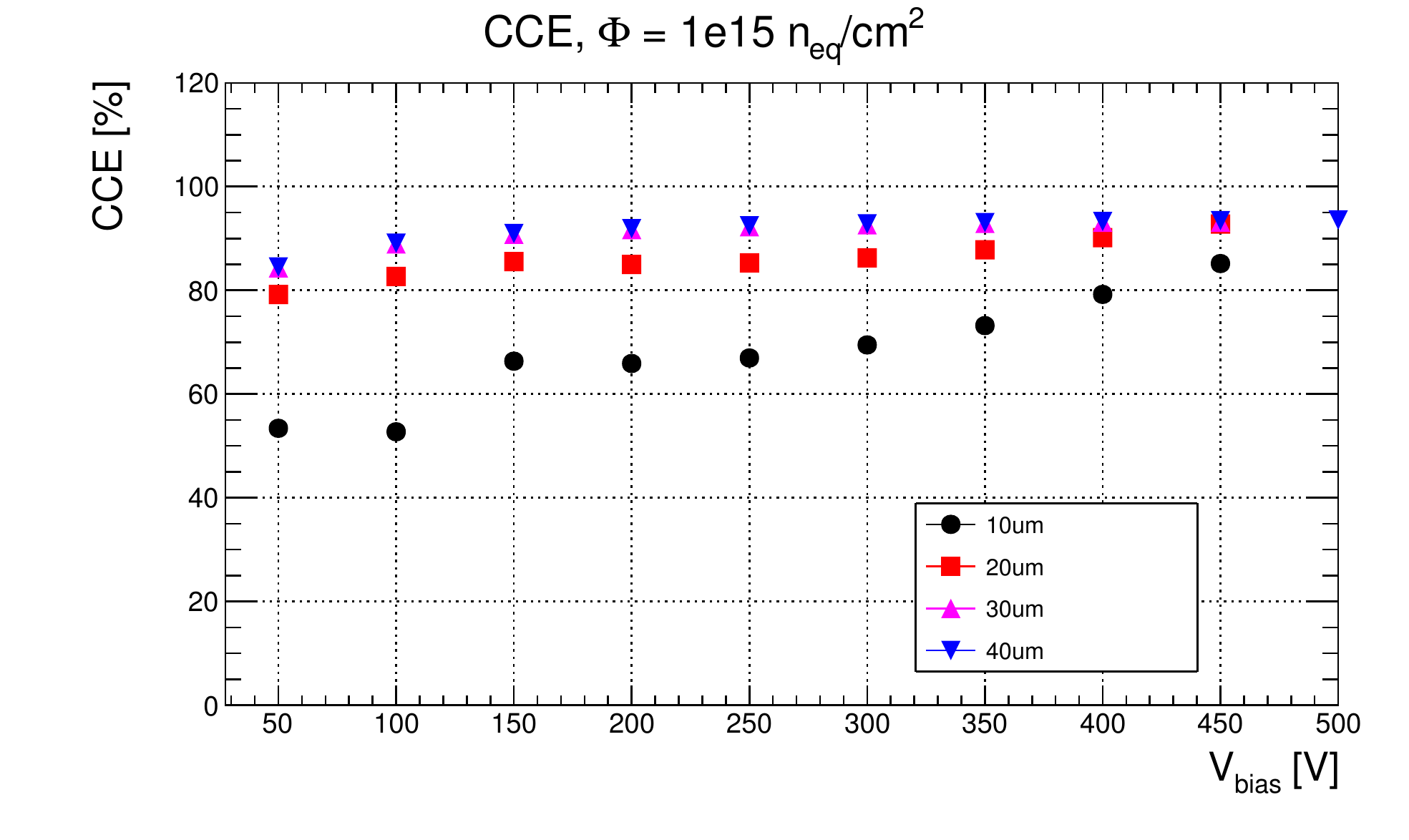}
\includegraphics[trim=1.2cm .6cm .3cm 1.1cm, clip, width=0.49\textwidth]{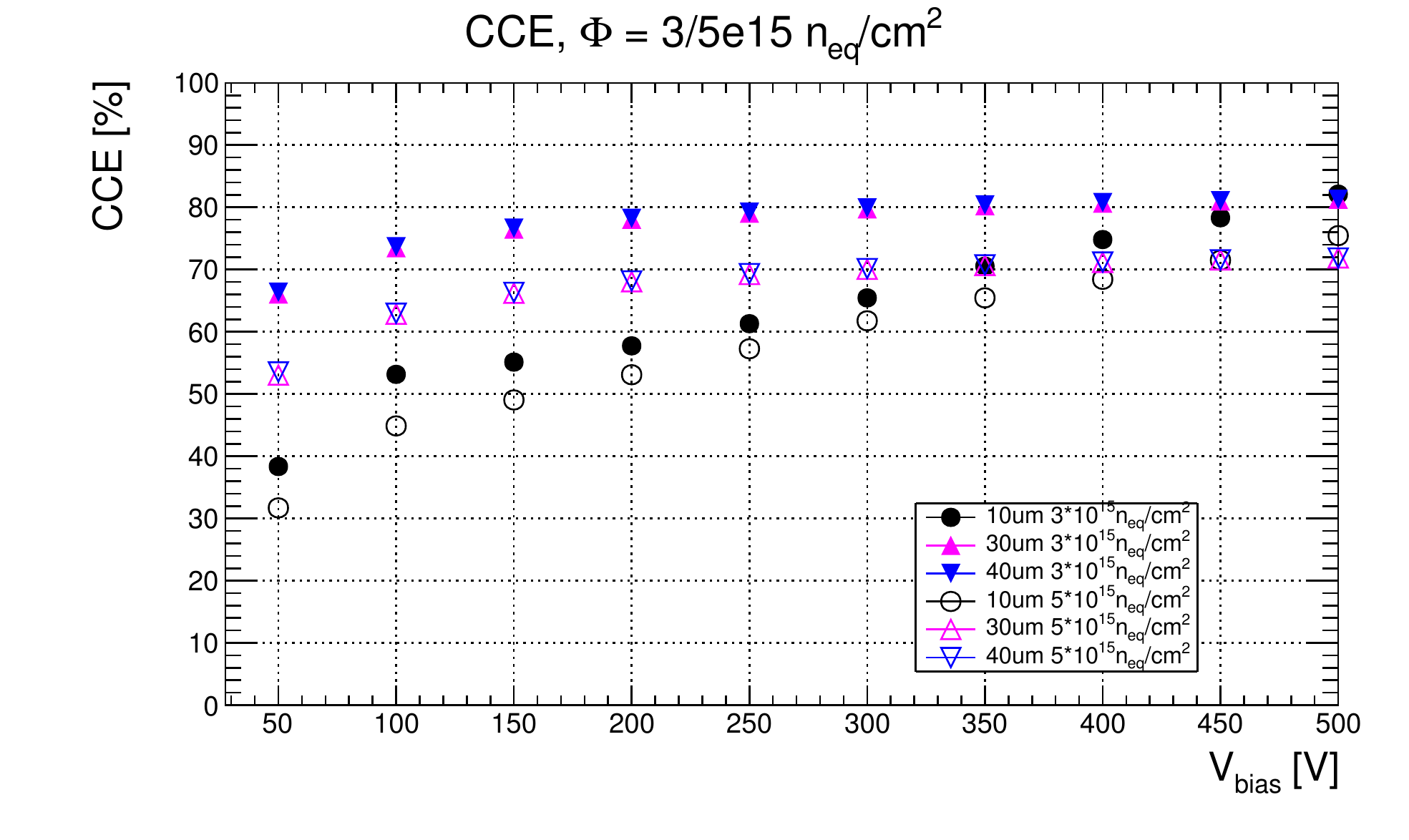}
\caption{\label{fig:CCEvsbiasirrad} Charge collection efficiency for different implant width as a function of bias voltage at $1\times10^{15}\,$n$_\text{eq}/$cm$^2$ (left) and 3 and $5\times10^{15}\,$n$_\text{eq}/$cm$^2$ (right). The particle traverses the sensor perpendicularly at 10$\,\mu$m/10$\,\mu$m, with 0/0 being the center of the four pixel region.}
\end{figure}

Figure \ref{fig:CCEvsbiasirrad} illustrates the CCE after irradiation. The left plot is for a fluence of $10^{15}\,$n$_\text{eq}/$cm$^2$. There is a difference in CCE between the two larger ($30\,\mu$m and $40\,\mu$m) and the two smaller ($20\,\mu$m and $10\,\mu$m) implant widths which is most pronounced for low voltages and decreases towards higher voltages. For irradiated structures, less charge is collected by smaller implants. Qualitatively, this does not depend on the fluence. At 3 and $5\times10^{15}\,$n$_\text{eq}/$cm$^2$, displayed on the right side of Fig.~\ref{fig:CCEvsbiasirrad}, the larger implants of $30\,\mu$m and $40\,\mu$m perform nearly equivalent at each fluence. The smaller implant of $10\,\mu$m needs a higher bias voltage of more than $400\,$V to reach the CCE of the larger implants.

\subsubsection{Spatially resolved CCE}
\label{section:spatialCCE}

\begin{figure}[tbph!]
\centering
\includegraphics[trim=1.2cm .6cm .3cm 1.1cm, clip, width=0.49\textwidth]{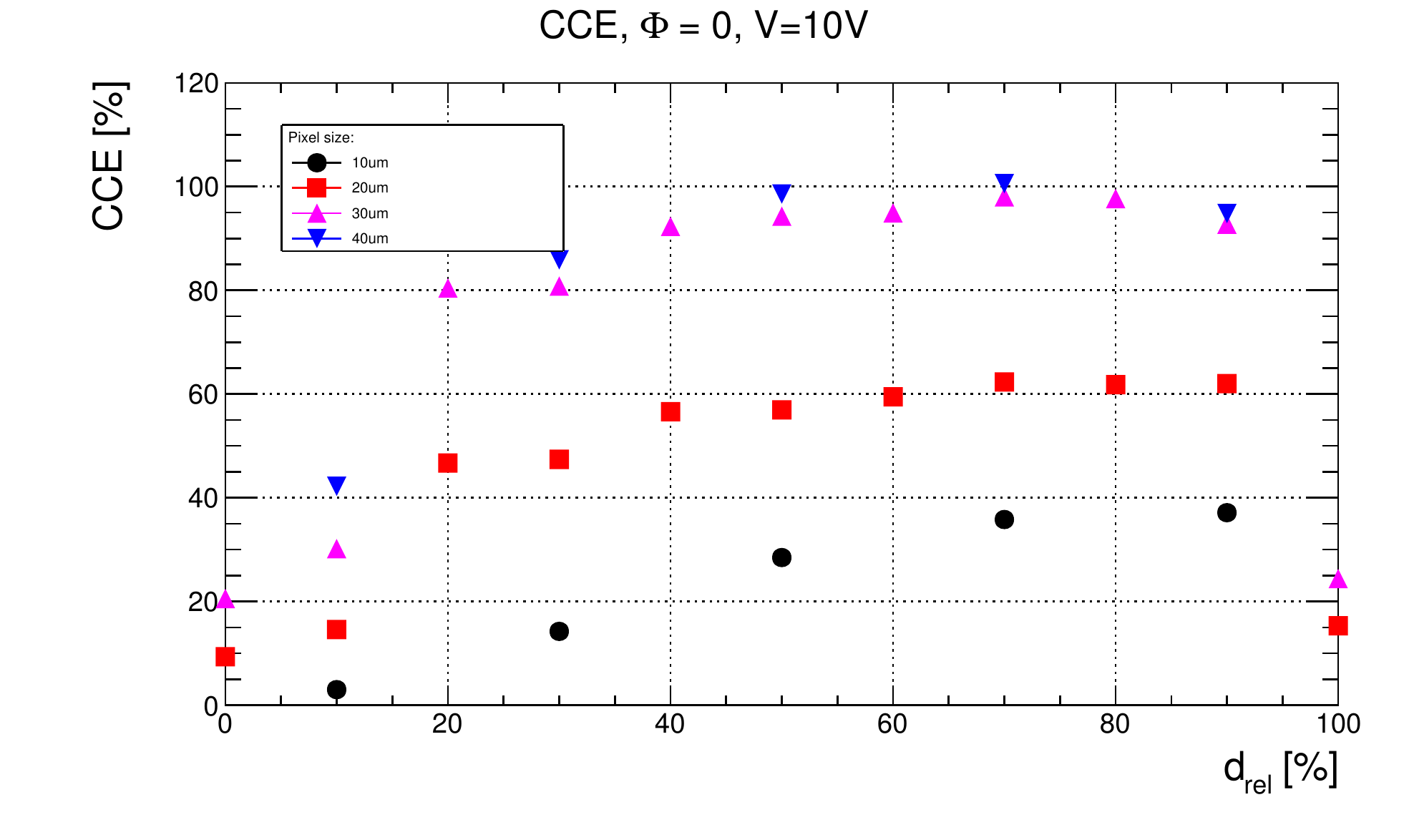}
\includegraphics[trim=1.2cm .6cm .3cm 1.1cm, clip, width=0.49\textwidth]{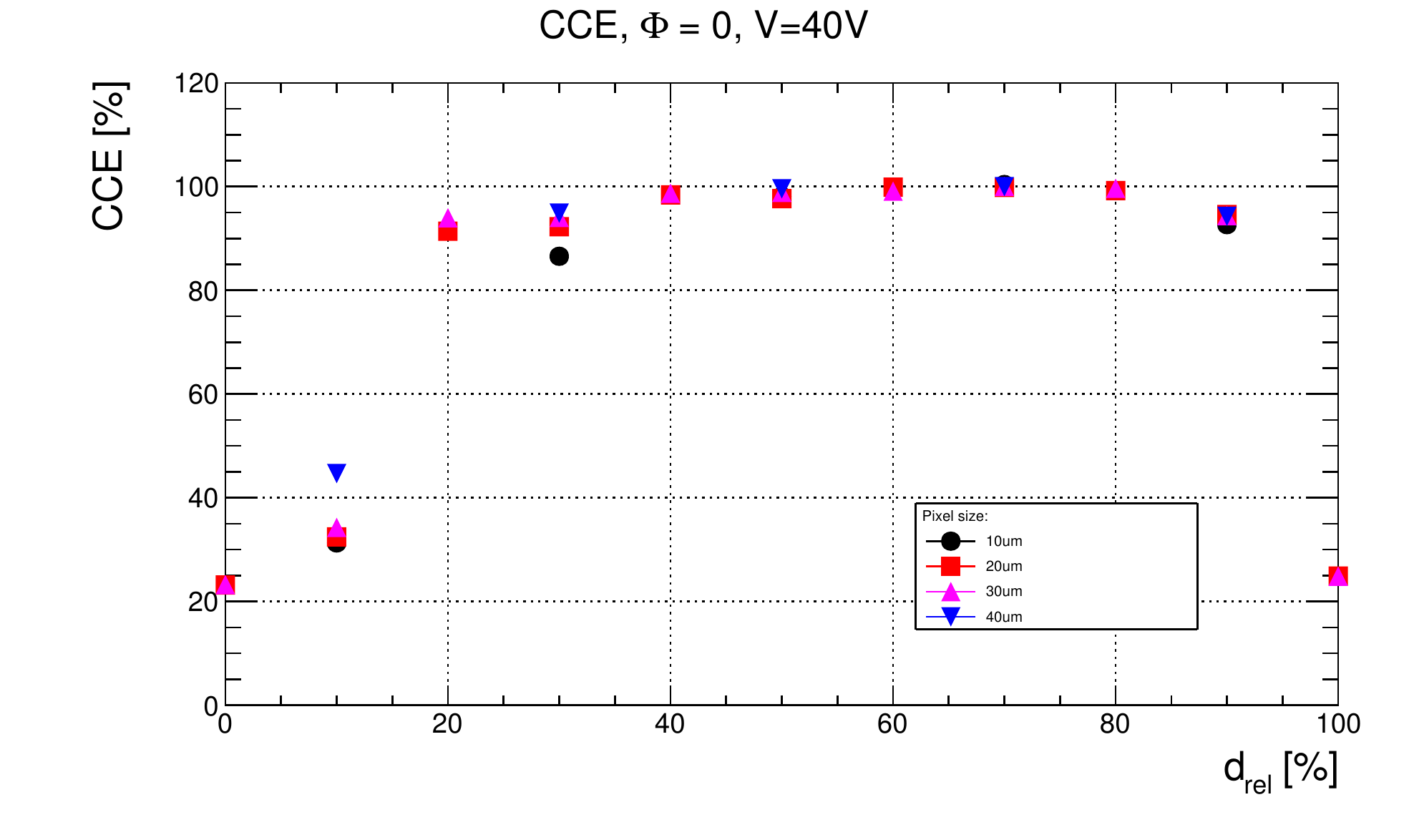}
\caption{\label{fig:CCEvsposition} Charge collection efficiency before irradiation as a function of relative position. The figure shows the CCE for different implant widths at a bias voltage of $10\,$V (left) and $40\,$V (right).}
\end{figure}

This section investigates the dependence of the collected charge on the point of charge injection. A perpendicular track penetrates the full thickness of the sensor while the impact point is varied across a diagonal starting in the center of the inter-four-pixel region (d$_\text{rel}=0\%$), ending in the center of the pixel under investigation (d$_\text{rel}=100\%$). The normalisation is again relative to the charge that is being collected by the largest implant (40$\,\mu$m) at $40\,$V. Figure \ref{fig:CCEvsposition} displays the behaviour at the two voltages determined in Sec. \ref{sec:ccevsvoltage} before and at depletion for not irradiated structures. The $10\,\mu$m and $20\,\mu$m wide pixel implants show a lower CCE across the whole diagonal at $10\,$V, agreeing with the hypothesis of not being depleted. Once depleted (right figure), the CCE is equal for all implant widths across most of the diagonal. Only the part (d$_\text{rel}\leq 30\%$) particularly prone to charge sharing in the region between the four pixels reveals a slightly better performance of the larger implants. The results at d$_\text{rel}=100\%$ deviate from the others since there is the corner of the structure at this position. Consequently, approximately three quarter of the charge is shared to the virtual continuation of the structure.

\section{Testbeam studies with annealed pixel modules}

The microscopic defects in the silicon crystal lattice caused by NIEL damage vary with time. They undergo an annealing process \cite{mollthesis} which leads to the deactivation of acceptor-like traps on a short timescale, causing a decreased full depletion voltage in p-type bulk. However, on a longer timescale, more acceptor-like traps are created than deactivated. The full depletion voltage after long annealing periods exceeds the full depletion voltage before annealing. The timescales of these beneficial and reverse annealing effects strongly depend on the ambient temperature. Both processes are strongly suppressed at the typical operational temperatures well below $0^\circ$C. Yet, the timescale of beneficial annealing is in the order of one week and the timescale of reverse annealing is in the order of several weeks at room temperature (RT). Periods in which the detector is kept at RT may occur during maintenance and upgrade phases. Since the innermost two layers of the ITk pixel detector will be replaced once during the ITk lifetime, it is possible that the outermost layer will have to be kept at RT during the replacement work for a long period of time in the order of several months.
\\
The effect of reverse annealing on the hit efficiency of highly irradiated modules is systematically investigated. Modules consisting of thin planar n-in-p pixel sensors interconnected to FE-I4 chips of the IBL detector with a pixel pitch of $50\times250\,\mu\mathrm{m}^2$ are used. These modules were tested at the testbeam facilities of CERN-SPS and DESY. A detailed description of the methodology of testbeam measurements can be found in \cite{Testbeam}.

\begin{figure}[tbph!]
\centering
\includegraphics[width=0.7\textwidth]{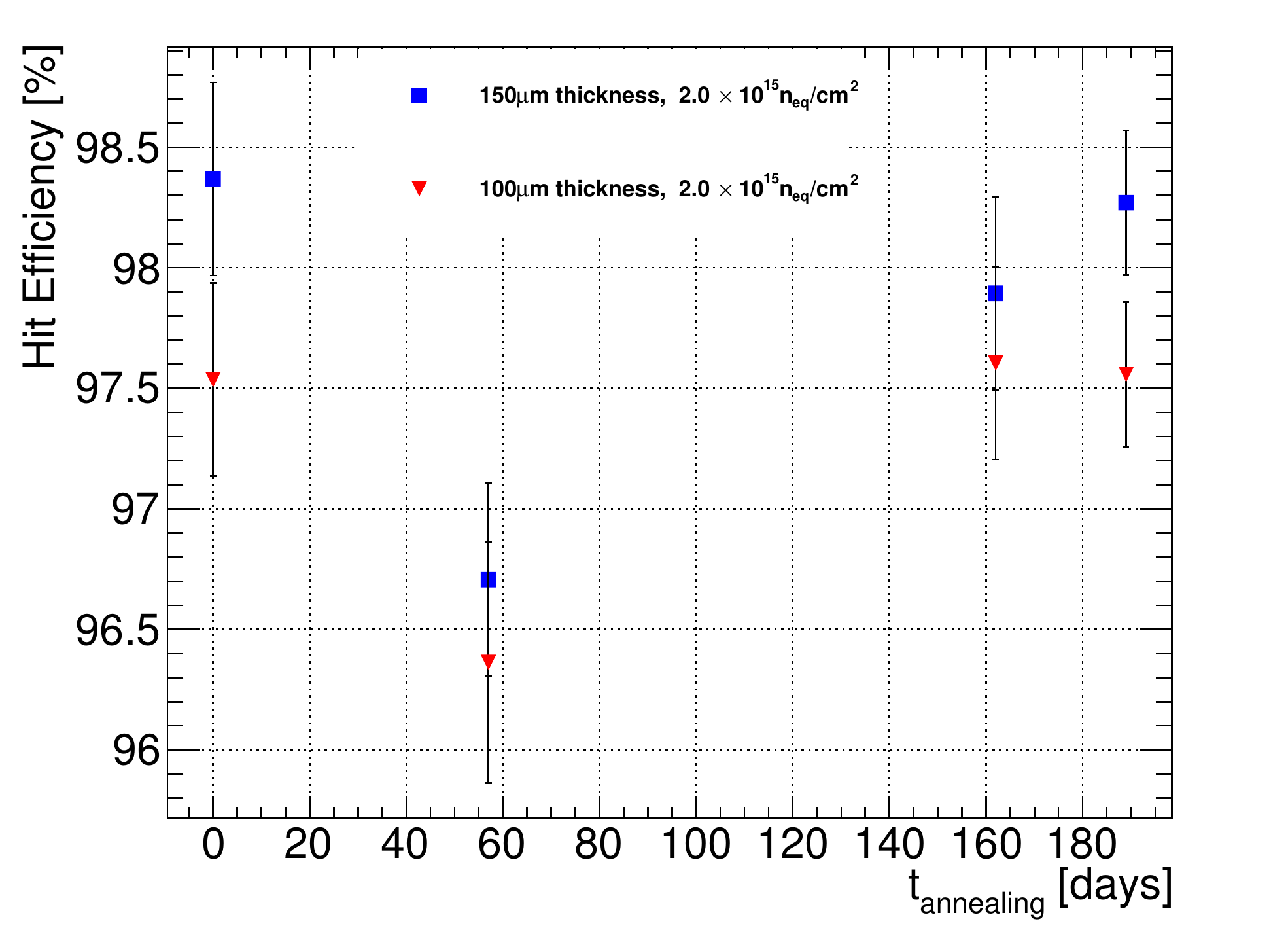}
\includegraphics[width=0.7\textwidth]{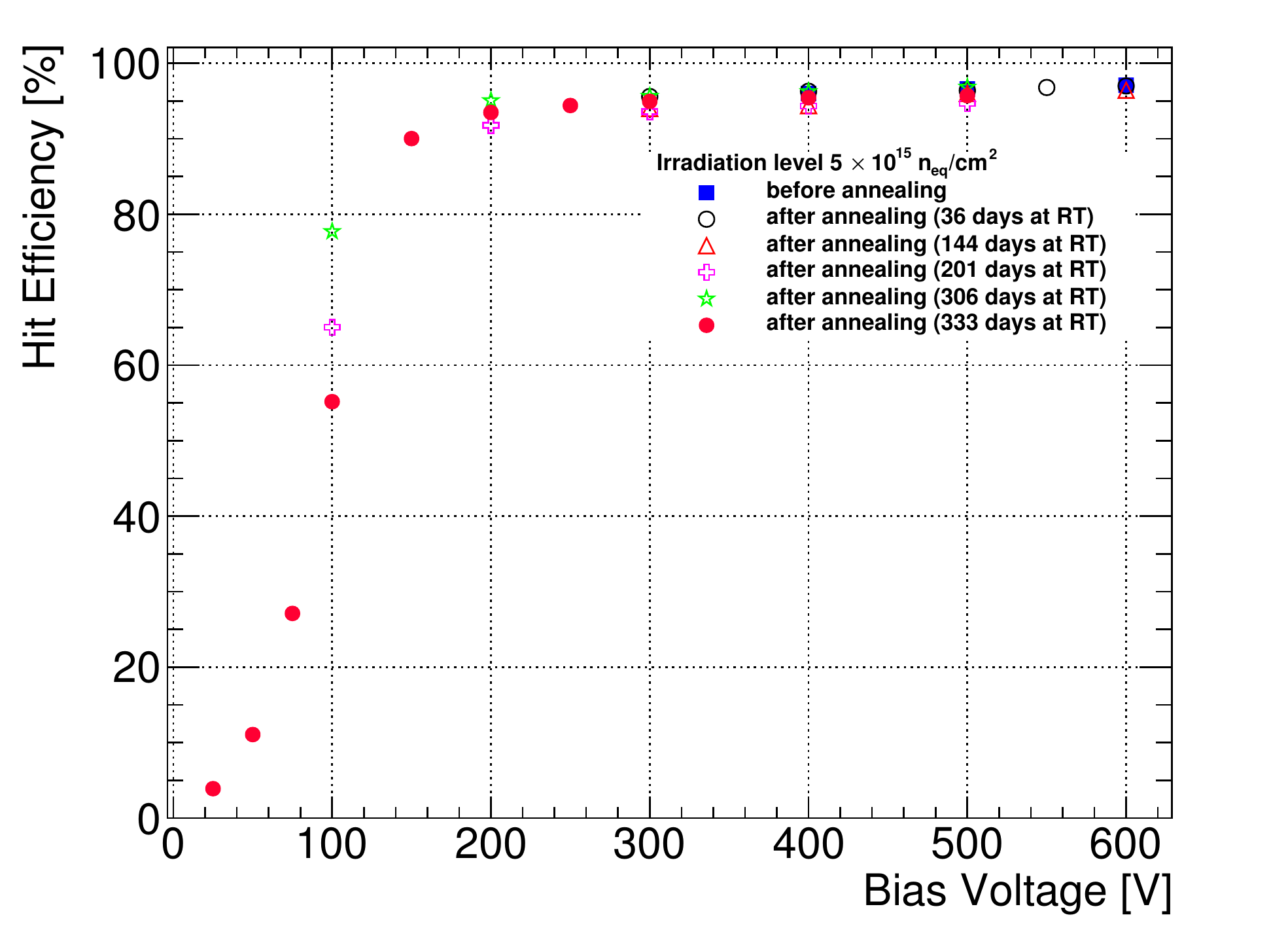}
\caption{\label{fig:annealingcomparison2e15} Hit efficiency as a function of annealing time at a fixed bias voltage of $500\,$V of the two modules irradiated at CERN (top) and the hit efficiency as a function of bias voltage at various stages of annealing of the module irradiated at KIT (bottom).}
\end{figure}

Three modules are presented here. All of them origin from the 2017 SOI4 production at HLL described in Sec. \ref{thinplanarpixelsensors}. Two modules were irradiated to a fluence of $2\times10^{15}\,$n$_\text{eq}/$cm$^2$ at the CERN irradiation facility using $24\,$GeV protons. The Gaussian beam profile used for the irradiation requires the selection of certain pixels matching a defined range of fluences to ensure comparability between different testbeams. The results include pixels with fluences of $(2.00\pm0.05)\times10^{15}\,$n$_\text{eq}/$cm$^2$. These two modules have sensors with thicknesses of $100\,\mu$m and $150\,\mu$m and were annealed for up to 189 days at RT. Figure \ref{fig:annealingcomparison2e15} (top) shows the hit efficiencies of the two modules as functions of annealing time at a fixed bias voltage of $500\,$V. The apparent difference between the second and all other measurements is likely caused by further systematic issues in this particular testbeam, as it generally had worse than usual resolution which is known to artificially cause bad efficiency. The following measurements yield hit efficiencies compatible with the first measurement, leading to the conclusion that annealing for up to 189 days at RT has no impact on the hit efficiency of modules featuring n-in-p sensors with thicknesses of $100\,\mu$m and $150\,\mu$m up to a fluence of $2\times10^{15}\,$n$_\text{eq}/$cm$^2$.

The third module was irradiated at KIT with protons of $25\,$MeV to a fluence of $5\times10^{15}\,$n$_\text{eq}/$cm$^2$. The sensor of this module has a thickness of $100\,\mu$m and was annealed for up to 333 days at RT. Figure \ref{fig:annealingcomparison2e15} (bottom) shows the hit efficiency as a function of bias voltage at various stages of annealing. For some of the measurements during the annealing time the hit efficiencies were measured only for large bias voltages. As a consequence, the lowest voltage at which the hit efficiency was measured at all times is $400\,$V. No significant change in efficiency over 333 days of annealing is observed. At a lower voltage of $100\,$V some variation is observed, but no clear trend can be identified yet. All modules will be further measured at larger annealing times.

\section{Conclusion and outlook}

The characteristics of small pixel cells addressing the requirements for the ITk pixel detector were studied using a TCAD simulation. Larger implants result in larger breakdown voltage and CCE. Not only larger implants do reach higher bias voltages before breaking down, but they also deplete earlier thus enabling full charge collection at lower voltages. Larger implants were also found to be able to collect charge more efficiently from the inter-pixel region, reducing charge sharing. On the other hand, the capacitance was found to be strongly increasing with the pixel size. The consequent drawback is the increase of the noise, counteracting the beneficial effect of the higher signal of the largest implant sizes. A quantitative analysis can only be done once the relation of input capacitance to noise is fully characterised for RD53A read-out chips.

The testbeam study of irradiated pixel modules using readout chips of the IBL employing thin sensors of the planned ITk revealed no detrimental impact of annealing up to 333 (189) days at RT on hit efficiency at a fluence of $5\times10^{15}$ ($2\times10^{15}$)$\,$n$_\text{eq}/$cm$^2$. More investigations at longer annealing times are planned. If all studied modules confirm this observation, fewer environmental constraints during the replacement of the innermost two layers after half of the ITk lifetime would be necessary.


\acknowledgments


This work has been partially performed in the framework of the CERN RD50 Collaboration. The authors thank F.~Ravotti for the irradiation of modules at the CERN IRRAD facility at CERN-PS and A.~Dierlamm for irradiation at KIT. Supported by the H2020 project AIDA-2020, GA no. 654168. The authors thank the EUTelescope, TBmon2 and EUDAQ software developer teams. The authors thank J{\"o}rn Schwandt from the University of Hamburg and Rainer Richter and Peter Lechner from HLL for many fruitful discussions and helpful suggestions.


\end{document}